\documentclass[letterpaper]{JHEP}
\usepackage[dvips]{graphicx}
\input epsf.tex
\def\be{\begin{equation}}
\def\ee{\end{equation}}
\def\bea{\begin{eqnarray}}
\def\eea{\end{eqnarray}}
\def\ba{\begin{array}}
\def\ea{\end{array}}

\def\part{\partial}

\preprint{SUGP-01/11-2\\
 hep-th/0111121}

\title{Matrix String Theory and the Myers Effect}
\author{Pedro J. Silva\thanks{E-mail address: {\tt psilva@physics.syr.edu}}
       \\Physics Department, Syracuse University, Syracuse, New York 13244}
\date{DTP-yy-nn}
\abstract{A new configuration of non-abelian D1-branes growing into D5-branes is found.
This time the effect is triggered by a non-trivial electric field on the world-volume of the D1-branes
and a constant RR 4-form potential. Based on the these configurations and other observations
regarding non-abilean effective actions, a new action for matrix string theory in non-trivial
backgrounds is conjectured. As an application we found that fundamental strings can
grow into Dp-branes, in particular by placing the strings in the background of a group of near horizon
D3-branes we found D5-branes. These types of configurations were found from the supergravity point
of view in previous works.}

\begin{document}

\newpage
\section{Introduction}

With the appearance of the web of dualities in string theory
a lot of attention has been paid to the non-perturbative aspects of
the theory. A new picture arises where the collection of all of
the supersymmetric string theories plus their dualities and d=11
supergravity form part of a whole that has come to be known as
M-theory. While the actual definition of this theory is a bit elusive,
we rely on some definitions that though are not
complete, still can give us an indication of what the theory should be.
In particular, the Matrix theory proposal \cite{bfss} has proved to be
very successful.

Matrix string theory \cite{dvv,bs,kro,cj1} is one of the outcomes of the different
dualities. Perhaps a simple way to define it is by looking at
Matrix theory with an extra compactified dimension. For example,
following Dijkgraaf et. al. \cite{dvv}, if we consider M-theory on a torus of radii
$A$ and $B$, by first reducing  on $A$  and then making an infinite boost on $B$
we get type IIA string theory on the discrete light cone (DLC)
with D0-brane particles. If, on the other hand, we reduce on $B$ first,
boost and then consider t-duality on $A$, we get (1+1) super Yang-Mills (SYM)
theory with fundamental string charge on the world-volume i.e.
the low energy theory of D1-branes.
Therefore, one finds that Matrix string theory is a non-perturbative
definition of string theory built in terms of a two dimensional SYM
theory and a collection of scalar fields in the adjoint representation
of the gauge group (see the original papers for an extended
discussion of this derivation). Although we have  discussed just
type IIA string theory, there are other constructions similar to the
one sketched before, where the other four superstring theories are
written in terms of two dimensional SYM theory\footnote{Actually for
the cases of type IIB and type I string theory there is an additional construction
in terms of a three dimensional theory \cite{cj1}}.

The Matrix string theory conjecture was  originally formulated on flat
backgrounds. Lately, using some techniques developed by Taylor and
Raamsdonk \cite{tvr1} a generalization for closed strings on non-trivial
weak backgrounds has appeared \cite{sch}. Also, by demanding
consistency of the different dualities involved in the above
process, a transformation law for the world-volume fields of the
D1-brane under s-duality was postulated.

In that very same article \cite{sch}, the possibility of non-abelian
configurations of fundamental strings was pointed out. In
particular the appearance of a Myers-like effect was computed
explicitly. By now the Myers effect \cite{mye1} is a well known
phenomenon where $N$ D-branes adopt a non-abelian configuration
that can be understood as a higher dimensional abelian D-brane.
These configurations come about as the result of new interaction
terms that appear in the non-abelian effective actions\footnote{
There is a paper with M-theory versions of these effective
actions, and with fundamental string actions in terms of matrices,
although the connection with matrix string theory is not clear
\cite{yol}.}.

Up to now (to the best of our knowledge), there is no construction
of the Myers effect in terms of fundamental strings. Nevertheless, given
the form of the effective action derived in \cite{sch},
this is a natural thing to expect. Actually, there is a paper of J. M. Caminos
et. al. \cite{alf} where the existence of fundamental strings forming
exotic configurations of D-branes is suggested. In that case
the construction resembles a baryon vertex configuration where in
the background of $M$ near horizon Dp-branes they find
a supersymmetric probe D(8-p)-brane, made out of $m$ fundamental strings.
 In the above paper the construction is based on the dual picture of
the higher dimensional abelian D(8-p)-brane. Actually, there are computations
 of Dp-branes collapsing into fundamental strings \cite{mal} and fundamental strings
 blowing up into Dp-branes \cite{rem} in terms of abelian Born-Infeld actions.

The purpose of this paper is two-fold: first in section 2, inspired by the
above results in M-theory we will show a new type of Myers effect
related to the presence of quantized electric flux on the
world-volume of the brane. Second, in section 3, based on the new s-duality rules
postulated in \cite{sch} we propose a close form for the Lagrangian of matrix
string theory on non-trivial backgrounds. This Lagrangian is none other than the
Myers non abelian Lagrangian for the D1-brane after an appropriate chain of
dualities is taken. Then we check that at lowest order this reproduces the matrix
string theory Lagrangian and agrees with the weak coupling calculations
of \cite{sch}. In section 4, using Matrix string theory on
non-trivial backgrounds we obtain the configurations of the type
described in \cite{alf}, this time from the point of view of non-abelian
fundamental strings. The final section is a conclusion and summary of the
results presented.

\section{D1-brane/D5-brane and the non-trivial electric field}

Although in the literature there are  some examples of p-branes
growing into (p+4)-branes \cite{mye2,bcp,dp}, these configurations
differ in their nature from the one we will present here,
since in our case it is the coupling with the world-volume
gauge field what triggers the non-abelian configurations.
Here, for simplicity we will work on the $p=1$ case only.

Our starting point is the low energy action for N D-strings with
non-trivial world-volume electric gauge field $F$. In the background we have a
trivial dilaton $\phi$, a flat metric $G$ and zero B-field. For the Ramond sector we
include a 4-form potential $C^{[4]}$. The Born Infeld action is,
\bea
S_{BI}=-T_1 \int d\tau d\sigma \, STr\left( e^{-\phi}
\sqrt{-\left( P\left[G_{ab}+G_{ai}(Q^{-1}-\delta)^{ij}G_{jb}
\right]\right) \, det(Q^i{}_j)} \right)
\label{eq:1}
\eea
with
\be
Q^i{}_j\equiv\delta^i{}_j + i\lambda\,[\Phi^i,\Phi^k]\,G_{kj}\ .
\ee
In writing (\ref{eq:1}) we have used a number of conventions taken from
Myers \cite{mye1}:
\begin{itemize}
\item{}
Indices to be pulled-back to the world-volume (see below)
have been labelled by $a$.  For other indices,
the symbol $A$ takes values in the full set of space-time coordinates while
$i$ labels only directions perpendicular to the center of mass world-volume.
\item{}
The parameter $\lambda$ is equal to $2\pi l_s^2$.
\item{}
The center of mass degree of freedom does not
decouple\footnote{We thanks D. Sorokin for useful comments on
this. Actually the center of mass degrees of freedom seems to play
an important role in the search for a covariant version of these
effective actions \cite{dso}.}, but they will be not relevant for
our discussion as we will consider static configurations
independent of the space-like world-volume direction. The fields
$\Phi^i$ thus take values in the adjoint representation of SU(N).
As a result, the fields satisfy $Tr \Phi^i=0$ and form a
non-abelian generalization of the coordinates specifying the
displacement of the branes from the center of mass. These
coordinates have been normalized to have dimensions of
$(length)^{-1}$ by multiplication by $\lambda^{-1}$.
\end{itemize}
The rest of the action is given by the
non-abelian Chern-Simons term. This term involves the non-abelian
`pullback' $P$ of various covariant tensors to the world-volume of
the D1-brane. We will use the static gauge $x^0=\tau,x^1=\sigma,x^i = \lambda \Phi^i$
for a coordinate $x$ with origin at the D1-brane
center of mass. The symbol $STr$ will be used to denote a trace
over the $SU(N)$ indices with a complete symmetrization over the
non-abelian objects in each term.  In this way, the Chern-Simons
term may be compactly written as
\bea
S_{CS}=\mu_1\int d\tau d\sigma
STr\left( P\left[ e^{i\lambda\,i_\Phi i_\Phi}(C^{(4)})
\right] \, e^{\lambda F}\right)\ ,
\eea
where the symbol $i_\Phi$ is a non-abelian generalization of the
interior product with the coordinates $\Phi^i$,
\be
i_\Phi \left(\frac{1}{2}C_{AB}dX^AdX^A\right) = \Phi^iC_{iB}dX^B.
\ee

If we restrict our study to static configurations involving five nontrivial
scalars $\Phi^i$, ${i=1,..,5}$, the above action gives the following potential:
\bea
&&V(\Phi,F_{[2]},F_{[5]})=\lambda^2\left[{1\over 2}(\partial\Phi)^2+
{1\over 4}\Phi^{ij}\Phi_{ji}+{1\over 4}F^2_{[2]}\right]+ \nonumber \\
&&+ \lambda^4\left[ {1\over 4}\partial\Phi^i\partial \Phi_i\Phi^{jk}\Phi_{kj}-
{1\over 2}\partial\Phi^i\partial \Phi^j\Phi_{jk}\Phi^k_{\;\;i}-
{1\over 8}\Phi^{ij}\Phi_{jk}\Phi^{kl}\Phi_{li}+\right. \nonumber \\
&&\hspace{4cm}\left. +{1\over16}F^2_{[2]}\Phi^{ij}\Phi_{ji}+
{1\over 10}\Phi^i\Phi^j\Phi^k\Phi^l\Phi^mF_{mlkji}F_{[2]}\right],
\label{eq:2}
\eea
where we have used the convention that $\Phi^{ij}=[\Phi^i,\Phi^j]$.
We will consider the ansatz corresponding to the fuzzy four sphere,
\bea
\Phi^i=\pm a\,G^i\,\,,\,\,i=1,...,5
\eea
where $a$ is a positive constant and $G^i$ are the five matrices of Castelino et. al. \cite{clt},
a symmetric tensorial product of $n$ factors, each one made out of the Euclidean gamma matrices
(in the $4$-dimensional representation) with $n-1$ identity matrices.
These matrices are of dimension $N=(n+1)(n+2)(n+3)/6$.
For the RR five form we use the ansatz
\be
F_{ijklm}=-f\lambda^{-1/2}\epsilon_{ijklm},
\ee
where $f$ represents the strength of the RR field.

\begin{figure}
\begin{center}
\includegraphics[angle=-90,scale=0.5]{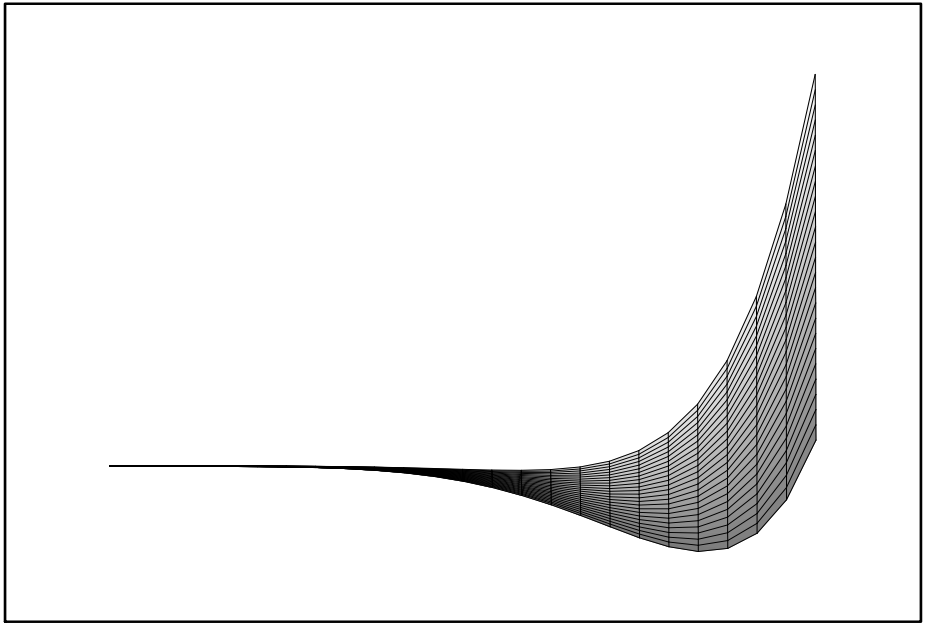}
\vspace{-1.0in}
\caption{{\it Potential for the D-string with non-trivial world-volume electric field
F and background RR 5-form field strength f. In the figure we show the potential as a
function of the radius $a$ and $f$, where we have considered $F_{01}=10, \lambda=1$. The line of
the bottom correspond to $f=-10$ the line on the top corresponds to $f=0$.}}
\label{fig1}
\end{center}
\end{figure}

Instead of solving the full equations of motion resulting from the previous action,
because all the matrix products simplify in the field equations, it is sufficient to substitute
the ansatz into the potential obtaining,
\bea
V =\lambda^2\left[4ca^4-{1\over 2}F^2_{01}\right]
+ \lambda^4\left[ 8c(c-8)a^8-cF^2_{01}a^4 - {4(n+2)c F_{01}fa^5\over 5\lambda^{1/2}}\right]
\eea
where $c$ is the Cassimir of the matrices $G^i$, given by the expression $c=n(n+4)$.
This potential clearly has a global minimum at a given value of $a$ that we will call $a_0$.
It depends on the value of the electric field $F$, the strength of the background RR field and $n$.
The actual analytical solution is not very enlightening, but here, in figure \ref{fig1}
we show a plot of the potential.
In particular if $F_{01}=2/\lambda$, the expression for $a_0$ simplifies, giving
\bea
a=\lambda^{-1/2}\left((n+2)f\over 4(n(n+4)-8)\right)^{1/3}.
\eea
Note nevertheless, that the physical radius is given by
\bea
\hat{R}=\lambda \sqrt{Tr(\Phi^i\Phi_i)/N}=\lambda\sqrt{c}a.
\eea
Hence,
\be
\hat{R}=\lambda^{1/2}\left((n+2)(n(n+4))^{3/2}f\over 4(n(n+4)-8)\right)^{1/3},
\ee
where we can see that the radius increases with the number of D1-branes and the strength of the RR field.
Also, note that in order to achieve this effect we need a nontrivial electric field on the D1-brane,
so we have fundamental strings diluted into the D1-brane as a requirement.

This configuration of non-abelian D-strings can be understood as a six
dimensional hypersurface. In fact, we can check that it correspond to a D5-brane configuration.
Following Myers et. al. \cite{mye2}, we consider the coupling of the non-abelian D1-branes
to a $C_{[6]}$ RR form
\bea
{\mu_1\over2}\int{Str\left(P[\lambda^2(i_\Phi i_\Phi)^2]C_{[6]}\right)}.
\eea
Using the fact that $C_{[6]}$ must have support on the fuzzy tunnel we write
\bea
C_{\tau\sigma ijkl}=C_{\tau\sigma\theta_1\theta_2\theta_3\varphi}{\epsilon_{ijklm}\over r}x^m
\eea
and hence we get,
\be
{\mu_1\over2}\int{d\tau d\sigma Str\left(\lambda^3C_{\tau\sigma\theta_1\theta_2\theta_3\varphi}
{\Phi^i\Phi^j\Phi^k\Phi^l\Phi^m\epsilon_{ijklm}\over R}\right)}.
\ee
Therefore, after using the fuzzy solution and $\mu_1=4\pi^2\mu_5\lambda^2$ and $\Omega_4=8\pi^2/3$ we get,
\be
\mu_5 {6N(n+2)\over c^{2/3}}\int{d\tau d\sigma \Omega_4R^4C_{\tau\sigma\theta_1\theta_2\theta_3\varphi}}.
\ee
which in the limit of large $n$ takes the form of the coupling of $n$ D5-branes
\footnote{Where the world-volume of the n D5-branes is taken along the $\tau,\sigma$ and the $S^4$
directions, and we take an average on the $S^4$.}.
\be
\mu_5 n\int{d\tau d\sigma \Omega_4R^4C_{\tau\sigma\theta_1\theta_2\theta_3\varphi}}.
\ee
Also, as another check we can calculate the form of the coupling of the D5-branes and the RR
potential $C^{(4)}$. Let us write the flat metric in coordinates adapted for this particular
configuration i.e.
\bea
ds^2=-d\tau^2+d\sigma^2+dr^2+r^2d\Omega^2_4 + dX^2 , \nonumber
\eea
where $d\Omega^2_4$  is the metric on a unit four dimensional sphere with angles
$\theta_1,\theta_2,\theta_3,\varphi$ and $dX^2$ are the remaining unimportant directions.
As we already said, the D5-brane expands along the coordinates ($\tau,\sigma,\theta_1,\theta_2,\theta_3,\varphi$).
 In this coordinates the RR 5-form and its potential 4-form $C^{(4)}$ are given by,
\bea
F^{(5)}_{ijklm}={f \over \lambda^{1/2}}\epsilon_{ijklm}\;\;,\;\; C^{(4)}_{\theta_1\theta_2\theta_3 \varphi}=
{fr^5\over 5\lambda^{1/2}}sin(\theta_1)^3sin(\theta_2)^2sin(\theta_3)
\eea
The coupling of one D5-brane with this non-trivial backgrounds is given by,
\bea
{\mu_5}\int{ d\xi^6 P\;[C{(4)}\wedge \lambda F^{(2)}]}.
\eea
where $F^{(2)}$ is the world-volume gauge field strength. Therefore, substituting
the explicit form of $C^{(4)}$ and integrating on $S^4$ we get (up to numerical factors),
\bea
{-1\over \lambda^{5/2}g}\int{ d\tau d\sigma \epsilon^{ab}F_{ab}fr^5}.
\eea
Then, using that the radius $r$ is given by the expression
\be
r=\sqrt{Tr[\sum_i \Phi^2]/N}\propto a\lambda,
\ee
we get
\bea
{-\lambda^{5/2} \over g}\int{ d\tau d\sigma \epsilon^{ab}F_{ab}fa^5}.
\eea
which is precisely the form of the corresponding coupling for the non-abelian D1-brane.

\section{Matrix string and Non-abelian D1-branes}

In the previous section we saw how $N$ D-strings polarized into higher dimensional objects.
This calculation can be translated into F1-strings polarizing into other higher dimensional objects
 if we use the appropriate s-duality transformation on the D-string world-volume fields.
 But what would be the correct interpretation of this non-abelian theory in the weak coupling string limit?.
A possible answer comes from matrix string theory. This is a supersymmetric gauge theory
that contains DLC string theory and has extra degrees of freedom representing non-perturbative
objects of string theory. Also, it is a second quantized theory as it is built from many strings.

We know that by means of different dualities the five superstring theories are
described in the neighborhood of a 1+1 dimensional orbifold conformal field theory.
In this language the strings are free in the conformal field theory limit, representing
DLCQ string theory. The interactions between the strings are turned on by operators
describing the splitting and joining of fundamental strings. These operators deform the
theory away from the conformal fixed point.

To further clarify these ideas, let us follow Krogh's  sketch of the derivation for the case of
type I string theory \cite{kro}. Consider type I strings in the DLC frame with no Wilson lines,
string mass $m_I$, string coupling $g_I$ and a null compact direction of radius $R_I$ (where we
identify the null coordinate as $x^- \approx x^- + R_I$). We then perform an s-duality
transformation to heterotic strings with string mass $m_H$, string coupling $g_H$ and a
null compact direction of radius $R_H$, again in the DLC frame. The relations between the
heterotic and type I parameters are
\bea
m_H={m_I\over g_I^{1/2}}\;\;,\;\;g_H={1\over g_I}\;\;,\;\;R_H=R_I.
\eea
Using the relation between a null compactification and a space-like compactification a la
Seiberg-Sen \cite{ss}, we get heterotic string theory on a space-like circle of radius $R$
in the sector with momentum $N$, string mass $m$ and  string coupling $g$.
The relation between these two heterotic string theories is given by,
\be
m^2R=m_H^2R_H\;\;,\;\;g=g_H\;\;,\;\;R\rightarrow 0.
\ee
Next we perform a t-duality on $R$, so that the new constants of the string theory
$(m',g',R')$ are given by,
\be
m'=m\;\;,\;\;g'={g\over mR}\;\;,\;\;R'={1\over m^2R}.
\ee
finally, we perform a s-duality to obtain type I string with $N$ D1-strings, and constants
$(m'',g'',R'')$ given by the following expressions,
\be
m''={m'\over g'^{1/2}}\;\;,\;\;g''={1\over g'}\;\;,\;\;R''=R'.
\ee
In terms of the initial type I string theory and $R$ we get
\bea
&&m''=\left[{(m_I^2R_I)^3\over g_IR}\right]^{1/2}\rightarrow 0,\nonumber\\
&&g''=(g_Im_I^2R_IR)^{1/2}\rightarrow 0,\nonumber\\
&&R''={g_I\over m_I^2R_I}.
\eea
Therefore, we get the low energy theory of N D1-branes at weak coupling, where the gauge
coupling constant $g_{YM}$ is given by,
\be
g^2_{YM}=m_I^4R_I^2.
\ee
This is the 1+1 dimensional SYM theory with eight scalars in the adjoint representation of the
gauge group. This effective action is obtained by the dimensional reduction of $N=1$ supersymmetry Yang-Mills
theory in ten dimensions down to two dimensions.

The type IIB case is similar to the type I case (previous case), in that a duality relation is need between
string theories in the DLCF. In the type IIB case a t-duality is used to go from type IIB to type IIA
in the DLCF, and then by boosting, t-dualizing and s-dualizing we define the matrix
string theory. In this case we get
\bea
&&m''=\left[{m_B^2\over g_B^2R_BR}\right]^{1/4}\rightarrow 0,\nonumber\\
&&g''=\left({m_B^2R_BR\over g_B}\right)^{1/2}\rightarrow 0,\nonumber\\
&&R''=R_B,
\label{eq:b1}
\eea
where $(m_B,g_B,R_B)$ are the initial type IIB string parameters and $(m'',g'',R'')$ are the
final (also type IIB) string theory parameters. Again, we get a low energy weak coupled string
theory with N D1-branes. The gauge coupling constant $g_{YM}$ is given by,
\be
g^2_{YM}={m_B^2\over g_B^2}.
\label{eq:b2}
\ee
The heterotic and type IIA definitions are also very similar, where some care has to be
taken with the inclusion of Wilson lines for the heterotic cases \cite{kro}. In these three string
theories a nice picture emerges by reinterpreting the involved dualities in terms of
a flip of the two radii that define the corresponding compactified M-theory \cite{dvv}.

In order to obtain the relevant action for one of the five matrix string theories, we start
with the world-volume gauge theory of N D1-branes, and then go back along the chain of dualities
until we reach the desired DLCQ string theory. For example, consider first an s-duality
transformation on the D1-brane effective action, then a t-duality transformation and finally the boost
relations of Seiberg-Sen. As a result we get type IIA matrix string theory. This can be written as
\be
L^{IIA}_F\equiv B\circ T\circ S \;[L_{D1}].
\label{eq:3}
\ee
Other matrix string theories Lagrangians can be obtain by similar procedures. For example,
$L^{IIB}_F\equiv T\circ B\circ T\circ S \;[L_{D1}]$.

As we mentioned in the introduction, there are generalization of the matrix
string action to include weak backgrounds. This time the calculations are based on the
strong relation between matrix string and the matrix theory proposal.
In particular, previous works of Taylor and Van Raamsdonk \cite{tvr1} are used to support these results.
One of the positive outcomes of the above work is a proposal for the transformation of the D1-brane
world-volume fields under s-duality. Thus, based on these different proposals we are
able to actually construct the matrix actions using maps like the one in equation \ref{eq:3}.

It is important to note that recently Myers wrote a non-abelian action of N
Dp-branes in general backgrounds \cite{mye1} which is fully covariant under t-duality.
This action incorporates in the limit of weak backgrounds, all the couplings derived previously
by Taylor et. al. and also introduces some new ones. If we believe this effective action for
the D1-branes, we are forced to conjecture that:

{\it Matrix string theory is defined by this world-volume action plus the web of
dualities needed}.

Note that since the non-abelian D1-brane action proposed by Myers does not capture the full physics
of the infrared limit, we can only trust its expansion up to sixth-order in the field strength \cite{bai},
an this problems is inherit by the above conjecture for the matrix theory action.
Another technical problem comes from the chain of dualities, since it makes it difficult to give an
explicit closed form for the final Lagrangian. In particular, the t-duality map mixes RR
fields and NS fields. Nevertheless, we only have to use the Buscher rules \cite{bus} on the
supergravity background fields as t-duality (once we have s-dualized), leaves the world-volume
fields invariant.

For example, let us consider the type IIA case. Following equation \ref{eq:3}, we get that the
action for the Matrix string is given by two parts, the first corresponding to the original
Born-Infeld term of the D1-brane action,

\bea
\hbox{\large{S}}^{1}_{F1}={1\over\lambda}\int d\xi^2 Str \left\{
\sqrt{-det(P[\widetilde{E}+\widetilde{E}(\widetilde{Q}^{-1}-\delta)\widetilde{E}]+
\lambda e^{\widetilde{\phi}}\widetilde{g}F)det(\widetilde{Q})}\right\}
\label{eq:4}
\eea
where
\bea
&&\widetilde{E}_{AB}=\widetilde{G}_{AB}-e^{\widetilde{\phi}} \widetilde{C}^{(2)}_{AB},\nonumber \\
&&\widetilde{Q}^{\;i}_{\;j}=\delta^i_{\;j}+i\lambda[\Phi^i,\Phi^k]\widetilde{E}_{kj}
(\widetilde{g}e^{\widetilde{\phi}})^{-1},
\eea
and the tilde represents the t-dual transformation of the background fields. For example
the form of $\widetilde{C}_{AB}$ is
\bea
&\widetilde{C}_{AB}=\left( \begin{array}{cc}
C_{\alpha\beta y}+2C_{[\alpha }B_{\beta]y}-2C_yB_{y[\alpha}G_{\beta]y} &
\;\;\;C_\alpha-C_yG_{\alpha y}/G_{yy}\\
-C_\beta+C_yG_{\beta y}/G_{yy}& 0\end{array}\right)& ,
\eea
where the space-time index $A$ has been divided into the t-dualized direction $y$ and the other
directions $\alpha$.

The second part, corresponding to the original Chern-Simons term
of the D1-brane action is
\bea \hbox{\large{S}}^{2}_{F1}={1\over
\lambda}\int d\xi^2 STr\left\{ P\left[
e^{i\widetilde{g}^{-1}\lambda\,i_\Phi
i_\Phi}[(-\widetilde{B}+\widetilde{C}^{(4)})
e^{-\widetilde{C}^{(2)}}] \right] \, e^{\lambda
\widetilde{g}F}\right\}. \label{eq:5}
\eea
This action contains
the action of the matrix string theory of Dijkgraaf et. al.
\cite{dvv}, since by construction in trivial backgrounds the
D1-brane action of Myers gives the 1+1 SYM theory corresponding
the dimensional reduction on N=1 SYM in ten dimensions down to two
dimensions. Hence, by setting all of the background fields to be
trivial, we recover the standard form of type IIA matrix string
theory, \be \hbox{\large{S}}^{1}_{F1}=\lambda\int d\xi^2 Tr
\left\{ {\partial\Phi^2\over2}+{1\over
4g^2}[\Phi,\Phi]^2+{g^2\over 4}F^2 \right\}. \ee Also, all of the
linear couplings obtained by Schiappa \cite{sch} for the weak
field case, are derivable from the action of equation \ref{eq:4}
and \ref{eq:5}. It has been checked that the D1-branes linear
couplings found by Taylor et. al. are included in the non-abelian
action of Myers and the t-duality and s-duality relations are the
same as the ones used by Schiappa. Nevertheless, we have to keep
in mind that there are new couplings not considered before.

\section{Strings and non-commutative configurations}

Let us apply the above theoretical construction to a specific
case where the matrix string Lagrangian shows some of these new terms.
In particular, we will use the the configurations
studied by Camino et. al. \cite{alf}, where they predicted D(8-p)-branes
growing out of fundamental strings on the near horizon background
of $M$ Dp-branes. In their original analysis the whole construction
was based on the abelian D(8-p)-brane. This time we use the world volume
action of the fundamental strings and
therefore we will be able to construct this configurations from
the point of view of the non-commutative two dimensional action.
In particular, let us consider the case of p=3, where this time the matrix
string action should be placed in the near horizon geometry of $M$ D3-branes.
In this case the only non-trivial background fields are the RR five
form field strength
\bea
F_{[5]}=4R^4\Omega_5,
\eea
where $\Omega_5$ is the volume form on $S^5$, and the metric is
\bea
ds^2=\left({r\over R}\right)^2dx^2_{(1,3)}+\left({R\over r}\right)^2dr^2+R^2d\Omega^2_5,
\label{eq:6}
\eea
where $dx^2_{(1,3)}$ is the metric of a four dimensional Minkowski space-time and
$d\Omega^2_5$ is the metric of a unit five sphere $S^5$.
Here we are interested in a static configuration where the world-volume coordinates of the string
correspond to time and the radial directions $(t,r)$, on the background metric. The only
non-diagonal scalars $\Phi^i$, correspond to the directions along the five sphere $S^5$. This ansatz
is obviously based on the dual configuration of \cite{alf}. The matrix string Lagrangian
is given in terms of five non-diagonal $\Phi^i$, the Yang-Mills field living on the world-volume
${\cal F}$ and the RR five form field strength $F_{[5]}$. The form of the potential is
\bea
&&V=\lambda^2\left[{1\over 2}(\partial\Phi)^2+
{1\over 4g^2}\Phi^{ij}\Phi_{ji}+{g^2\over 4}{\cal F}^2_{[2]}\right]+ \nonumber \\
&&+ \lambda^4\left[ {1\over 4g^2}\partial\Phi^i\partial \Phi_i\Phi^{jk}\Phi_{kj}-
{1\over 2g^2}\partial\Phi^i\partial \Phi^j\Phi_{jk}\Phi^k_{\;\;i}-
{1\over 8g^4}\Phi^{ij}\Phi_{jk}\Phi^{kl}\Phi_{li}+\right. \nonumber \\
&&\hspace{4cm}\left. +{1\over16}{\cal F}^2_{[2]}\Phi^{ij}\Phi_{ji}+
{1\over 10g}\Phi^i\Phi^j\Phi^k\Phi^l\Phi^mF_{mlkji}{\cal F}_{[2]}\right],
\eea
\begin{figure}[t]
\begin{center}
\includegraphics[angle=-90,scale=0.5]{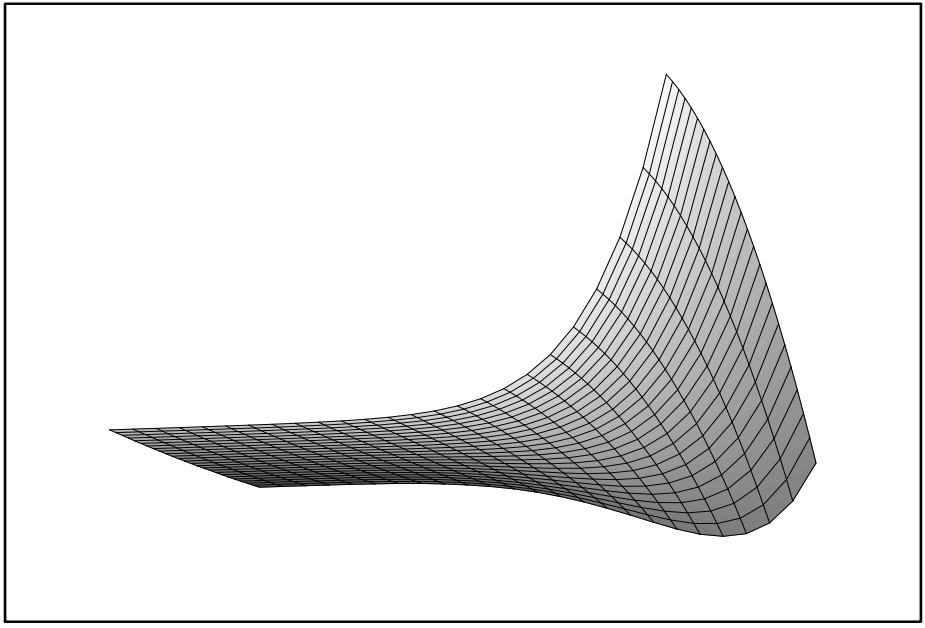}
\vspace{-1.0in} \caption{{\it Potential for the string stretching
along the $z$ coordinate in the near horizon geometry of a group
of D3-branes. In the figure we show the potential as a
function of $\rho$ and ${\cal F}$. The lower curve corresponds to
bigger values of ${\cal F}$, while the higher curve corresponds to
${\cal F}=0$.}}
\end{center}
\end{figure}
Note that the potential is very similar to the potential of
equation \ref{eq:2}. This is reasonable, since the fields involved
are s-dual invariant and the dilaton is trivial in this case.
Therefore, the t-duality involved is almost trivial. The real
difference comes in the relation between the string coupling
constant, the YM coupling constant and the DLC radius (see
equation \ref{eq:b1} and \ref{eq:b2}). Nevertheless, the metric is
non-trivial and the interpretation is different. Due to the form
of the metric, $S^5$ is not really accessible from the
embedding into $R^6$, since we will need an auxiliary extra
dimension, given a total of eleven dimensions. This implies a
technical problem if we want to use a fuzzy sphere ansatz, which
comes naturally in terms of cartesian coordinates. Nevertheless,
we know that the configuration we are looking for is a fuzzy $S^4$
in the $S^5$, therefore we rewrite the metric \ref{eq:6} as
\be
ds^2=\left({r\over R}\right)^2dx^2_{(1,3)}+\left({R\over
r}\right)^2\left( dz^2 + \delta_{ij}dx^idx^j\right),
\ee
where
$i=(1,...,5)$, $r^2=z^2+\delta_{ij}x^ix^j$, and $z$ coincides with
the space-like direction of the fundamental string. Next, using
the fuzzy $S^4$ ansatz $\Phi^i=aG^i$ and solving for this
specific metric and RR field, we get
\bea
&&V=\lambda^2\left[{c\over 2}\dot{a}^2+ 4c\left({Ra\over
rg^{1/2}}\right)^4-{g^2\over 2}{\cal F}^2_{01}\right]+
\lambda^4\left[ 4c(c-8)\left({Ra\over g^{1/2}r}\right)^4\dot{a}^2 + \right. \nonumber \\
&&\hspace{2cm}\left. + 8c(c-8)\left({Ra\over rg^{1/2}}\right)^8 - c{\cal F}^2_{01}\left({Ra\over r}\right)^4+
{4^2(n+2)c{\cal F}_{01}R^4za^5\over 5gr^6} \right],
\eea
where $\dot{a}$ corresponds to $\partial_za$. Let us consider the rescaling
\be
a=\left({r\over R}\right)\rho\;\;\Rightarrow\;\;r={zR\over \sqrt{R^2-\lambda^2c\rho^2}}.
\ee
Hence we get the following potential,
\bea
&&V=\lambda^2\left[{c\rho^2 \over 2(R^2-\lambda^2c\rho^2)}+
{4c\over g^2}\rho^4-{g^2\over 2}{\cal F}^2_{01}\right]+
\lambda^4\left[ {4c(c-8)\rho^6 \over g^2( R^2-\lambda^2c\rho^2)} + \right. \nonumber \\
&&\hspace{2cm}\left.  - c{\cal F}^2_{01}\rho^4+ {8c(c-8)\over g^2}\rho^8 +
{4^2(n+2)c{\cal F}_{01}\rho^5\sqrt{R^2-\lambda^2cg\rho^2}\over 5gR^2} \right].
\label{eq:7}
\eea
Again, there is a global minimum at $\rho_0$, defining the radius of the non-commutative brane.
In the figure we show the potential $V$ as a function of (${\cal F},\rho$), for fixed values
of ($R,N$).

Note that the radius of the fuzzy four-sphere is bounded by $R$ as it should be. Also,
the physical radius $\hat{R}$ is given by,
\be
\hat{R}=\lambda \sqrt{Tr(\Phi^i\Phi_i)/N}=\lambda \sqrt{c}{R \over r}a=\lambda \sqrt{c}\rho_0.
\label{eq:8}
\ee
and therefore is constant, independent of the position along the string.

The energy density of this configuration (along the "$r$" direction) corresponds to the value
of the potential of equation \ref{eq:7} evaluated at the minimum $\rho_0$.
Unfortunately, since we only have numerical solutions, an approximation is need to obtain
analytical results. Nevertheless, from fig. 2, we can see that the energy at $\rho_0$ is less
that for the case of zero radius.

The above non-commutative configuration of $N$ fundamental strings bits should be compared
to the configuration of D(8-p)-branes growing in the near horizon geometry of $M$ Dp-branes
for the case $p=3$ described in \cite{alf}. In order to compare these two dual pictures,
let us describe the dual configuration in more detail.

In the dual case, we have a stable configuration of $n$ D5-brane extending along the "$r$"
direction of the $AdS_5$ part of the metric, while the other four space-like directions are
partially wrapped on a $S^4$ embedded in $S^5$. The D5-brane contains an electric flux on
its world-volume, representing $m$ fundamental strings. The value of polar angle $\theta_m$
on the $S^5$ is given by
\bea
sin(\theta_m)cos(\theta_m)+\pi {m\over M}-\theta_m=0,
\eea
where $M$ is the number D3-branes that shape the geometry. This expression can be trusted
as long as the number of fundamental strings is much smaller than the number of D3-branes.
This is equivalent to considering only small $\theta_m$ angles. In this limit,
the value of theta is $\theta_m \approx \left({3\pi m\over 2M}\right)^{1/3}$. Therefore, the radius of the
four sphere, up to numerical factors is given by the expression
\bea
\hat{R}=Rsin(\theta_m)\approx {\lambda^{1/2}g^{1/4}m^{1/3}\over M^{1/12}}.
\label{eq:9}
\eea
In the same approximation, the form of the energy density (per unit length along the "$r$"
direction) is given by,
\bea
{\cal E}\approx {m\over \lambda} - {m^{5/3}\over \lambda M^{2/3}}.
\label{eq:10}
\eea
Note that the energy is always less than the energy of $m$ fundamental strings stretching
along the radial direction.

We can see that the shape and partial wrapping in both pictures (partonic and brane)
coincide. In both cases the physical radius of the four sphere $S^4$ doesn't depend on the
position in $AdS_5$ and this is a non-trivial check given that the background geometry is not flat.
To compare the radius $\hat{R}$, of equation \ref{eq:8}, to the dual expression \ref{eq:9}, we have to make
an approximation on the potential for small radius $\rho_0$. For example, in the case of large $N$,
but $\rho_0\lambda<<R$, we get that only the last two terms of \ref{eq:7} are relevant, giving
\bea
\rho_0\approx {1\over (nR)^{1/3}} \Longrightarrow \hat{R} \approx {\lambda^{1/2}g^{1/4}n^{2/3}\over M^{1/12}}.
\label{eq:11}
\eea
In the same approximation, the energy of the string bits gives up to numerical factors,
\bea
{\cal E}= {m\over \lambda}\left(1+ V(\rho_0)\right) \approx {m\over \lambda} - {n^{4/3}m\over \lambda M^{2/3}}.
\label{eq:12}
\eea
where we have included the overall contribution from the $STr$, that gives the number of "long strings" made out
of string bits (see for example \cite{dvv}).
In order to make a full comparison, we need to identify (in the string bits case), the number of "long strings"
that form the non-commutative geometry i.e. $m$ in the dual picture. Although this is a non-trivial task (where
a detail examination of the realization of the $G^i$ matrices is needed), here we note that if we take
$m \equiv n^2$, not only equations \ref{eq:9} and \ref{eq:11} match, but also equations \ref{eq:10}
and \ref{eq:12} (again up to numerical factors).

As a final comment, the configurations studied in \cite{alf}, proved to saturate a BPS inequality,
and therefore we expect the same here. Also, the partonic picture provides the mechanism that quantizes the
electromagnetic field in the corresponding dual picture, of the abelian D(8-p)-brane. This is precisely
the condition need to be imposed in the original paper, to obtain the stable D(8-p)-branes.

\section{Discussion}

In this work we conjecture that the natural extension of matrix string theory to non-trivial
backgrounds corresponds to the non-abilean action of $N$ D1-branes and the chain of dualities needed.
The fact that the Myers action \cite{mye1} is covariant under t-duality together with the approach
developed by Taylor and Van Raamsdonk \cite{tvr1}
and the work of Schiappa \cite{sch}, make this conjecture very solid and natural. Here, we followed
the work of Dijkgraaf et. al. \cite{dvv} and Krogh \cite{kro} to unravel the chain of dualities needed
to define the action of matrix string theory, from the effective action of D1-branes. Note that the final actions
of equations \ref{eq:5} and \ref{eq:5}, are only approximations, since it is known that action of Myers for
non-abelian D-branes do not captures all the physics in the general situation. Nevertheless abstract
relations like \ref{eq:3}, should be regard as exact statements.

Once we have these effective actions (one for each type of superstring theory), we can test new
physics by turning on only a few background fields.
In particular a Myers-like effect (a polarization of the string-bits), generates non-commutative
geometries that should have an interpretation as higher dimensional objects in M-theory. We should
recall that matrix string theory incorporates non-perturbative effects like D-brane physics.
The observation that the string bits should create higher dimensional objects is no new,
it was already presented in the work of Myers \cite{mye2} and Schiappa \cite{sch}, and from the supergravity perspective in the
work of J. M. Camino et. al. \cite{alf}.

From the point of view of D-brane physics, we found a new type of Myers effect, where D1-branes grow into
D5-branes. This configuration requires the presence of electric fields on the world-volume of the
D-brane, and therefore is also related to the presence of fundamental strings. These final configurations
resemble the fuzzy tunnel of Myers, although the physical interpretation is very different.

There is plenty of more work to be done, supersymmetry preserved
by the matrix string theory configurations should be checked
explicitly, although in the paper of J. M. Camino et. al.
\cite{alf}, these configurations were proved to satisfy a BPS
bound. Also the study of t-duality in the DLC is a tricky subject
and more care should be taken to define more rigourously the type
IIB matrix string theory.

\acknowledgments

The author would like to thank the organizers of the conference {\it Physics in the Pyrenees:
Strings, Branes and Field Theory} since this work is a direct consequence of this meeting, also
Alfonso Ramallo, Cesar Gomes, Yoland Lozano, Joel Rozowsky, Simeon Hellerman and Don Marolf
for useful discussions. Special thanks to Joel Rozowsky for proofreading
 the manuscript. This work was supported in part by NSF grant PHY-0098747 to Syracuse
University and by funds from Syracuse University.


\end{document}